\documentclass[aps,pra,reprint,amsmath,amssymb,superscriptaddress,nobibnotes]{revtex4-1}
\usepackage{graphicx,SIunits}
\usepackage{bm}     
\usepackage{dsfont}    
\usepackage{color}

\usepackage{multirow}
\usepackage{upgreek}
\usepackage{adjustbox}
\usepackage{tabularx}
\usepackage{array}
\usepackage{xcolor}
\newcolumntype{P}[1]{>{\centering\arraybackslash}p{#1}}
\usepackage{ulem}
\usepackage{csquotes} 
\usepackage{lineno}
\newcommand{\rom}[1]{\uppercase\expandafter{\romannumeral#1\relax}}

\usepackage{hyperref} 
\usepackage{float}

\begin{document}

\title{Robust and bright polarization-entangled photon sources exploiting non-critical phase matching without periodic poling}

\author{Ilhwan Kim}
\affiliation{Center for Quantum Technology, Korea Institute of Science and Technology (KIST), Seoul, 02792, Korea}
\affiliation{Department of Applied Physics, Kyung Hee University, Yongin-si 17104, Korea}

\author{Yosep Kim}
\affiliation{Department of Physics, Korea University, Seoul 02841, Korea}

\author{Yong-Su Kim}
\affiliation{Center for Quantum Technology, Korea Institute of Science and Technology (KIST), Seoul, 02792, Korea}
\affiliation{Division of Quantum Information, KIST School, Korea University of Science and Technology, Seoul 02792, Korea}

\author{Kwang Jo Lee}
\email{kjlee88@khu.ac.kr}
\affiliation{Department of Applied Physics, Kyung Hee University, Yongin-si 17104, Korea}

\author{Hyang-Tag Lim}
\email{hyangtag.lim@kist.re.kr}
\affiliation{Center for Quantum Technology, Korea Institute of Science and Technology (KIST), Seoul, 02792, Korea}
\affiliation{Division of Quantum Information, KIST School, Korea University of Science and Technology, Seoul 02792, Korea}

\date{\today} 

\begin{abstract}
Entangled photon sources are essential for quantum information applications, including quantum computation, quantum communication, and quantum metrology. Periodically poled (PP) crystals are commonly used to generate bright photon sources through quasi-phase matching. However, fabricating uniform micron-scale periodic structures poses significant technical difficulties, typically limiting the crystal thickness to less than a millimeter. Here, we adopt non-critical phase matching to produce a robust and bright polarization-entangled photon source based on a Sagnac interferometer. This method is tolerant of variations in pump incidence angles and temperature, and theoretically offers about a 2.5-fold brightness enhancement compared to quasi-phase matching. Additionally, the absence of periodic poling allows for a larger crystal cross-section. Using a bulk KTP crystal without a PP structure, we experimentally produce the four Bell states with a brightness of 25.1 kHz/mW, achieving purity, concurrence, and fidelity values close to 0.99. We believe our scheme will serve as a key building block for scalable and practical photonic quantum information applications.
\end{abstract}



\maketitle

Entangled photon pairs are essential for quantum communication~\cite{Yin2017_2,Jennewein2000,bulla2022}, quantum sensing~\cite{Giovannetti2004,Kim2024,Hong2021}, and quantum computing~\cite{Knill2001,brien2009,Kok2007}. Practical applications in various quantum information processing require bright, stable entangled photon sources. Several approaches have been proposed and demonstrated to generate entangled photon pairs, including the biexciton-exciton cascade in semiconductor quantum dots~\cite{Chen2024}, four-wave mixing with third-order nonlinearities in optical fibers~\cite{Garay2023} and waveguides~\cite{Grassani2016}, and spontaneous parametric down-conversion (SPDC) with second-order nonlinearities in diverse material platforms such as ferroelectric liquid crystals~\cite{Sultanov2024}, two-dimensional materials~\cite{Weissflog2023}, thin films~\cite{Okoth2019}, and bulk $\chi^{(2)}$ crystals~\cite{ Kwiat1999, Wang2016, Evans2010, Bruno2014, Chen2017_2, Kuklewicz2004, Kim2006, Fedrizzi2007, Yin2017, Weston2016}. Among these, SPDC in a second-order nonlinear crystal--a process in which a pump photon is  spontaneously converted into two lower-frequency signal and idler photons--has been widely used to prepare entangled photons due to its high entanglement quality, brightness, phase-matching flexibility, and low noise~\cite{Zhang2021}.


Polarization-entangled photon sources based on bulk beta-barium borate (BBO) crystals, which use birefringent phase matching (BPM), have been widely used for generating high-quality entangled photons~\cite{Kwiat1999,Wang2016}. However, the non-collinear propagation of the generated photon pairs in BPM leads to spatial walk-off, which constrains the length of the crystal, thereby limiting the brightness of the photon source~\cite{Yariv}. To address this issue, quasi-phase matching (QPM) has been introduced, utilizing a periodically poled (PP) structure in a nonlinear crystal~\cite{boyd}. The PP structure can eliminate spatial walk-off and relax the phase matching (PM) conditions by adjusting the period, making the collinear SPDC process possible in comparatively long crystals~\cite{Kuklewicz2004}. This results in high brightness, high spectral purity, and large tunability in the wavelengths of the generated photons. Since a polarization-entangled photon source using periodically poled potassium titanyl phosphate (PPKTP) based on a polarization Sagnac interferometer (PSI) was demonstrated~\cite{Kim2006}, it has been widely used to produce high-quality, bright, and stable polarization-entangled photon sources for the last two decades~\cite{Kim2024, Fedrizzi2007, Yin2017}.


Non-critical phase matching (NCPM)~\cite{Dmitriev}, a collinear SPDC process using a bulk nonlinear crystal without a PP structure, has the potential to improve QPM-based PSI for generating polarization-entangled photons. In QPM, the effective nonlinearity in a crystal decreases based on the order of grating vectors due to the PP structure~\cite{boyd}, leading to a reduction in brightness, whereas NCPM does not. Moreover, creating a PP structure within a nonlinear crystal requires sophisticated fabrication techniques, and the thickness of a QPM crystal is limited by the electrical conductivity of the materials~\cite{Shur2016}. In contrast, NCPM offers a broader beam window for nonlinear crystals, as it imposes no constraints on crystal thickness. This enables the generation of higher-dimensional entangled states in the spatial domain by employing multiple pump beams. For instance, the limited beam window of PP crystals restricts the number of pump beams to a few~\cite{Hu2021, Slussarenko2022}. However, it has been demonstrated that crystals without PP structures, which have a larger beam window (up to millimeter scale or beyond), can generate significantly higher-dimensional states~\cite{Hu2020, Li2020}.

In this work, we theoretically investigated and experimentally demonstrated the generation of a high-quality, ultra-bright polarization-entangled photon source using Type-\rom{2} NCPM in a bulk KTP crystal without a PP structure. We provided detailed theoretical calculations for NCPM in a KTP crystal and showed that the NCPM approach with KTP offers several advantages compared to QPM and BPM. We then constructed a polarization-entangled photon source based on a PSI and prepared four types of polarization-entangled Bell states. The generated Bell states exhibit high brightness, exceeding 25 kHz/mW, and high quality, with the concurrence and the state fidelity values approaching 0.99. We believe that our robust and ultra-bright entangled state has the potential for a wide range of applications in photonic quantum information processing.

\begin{figure}[t]
\centering\includegraphics[scale=0.7]{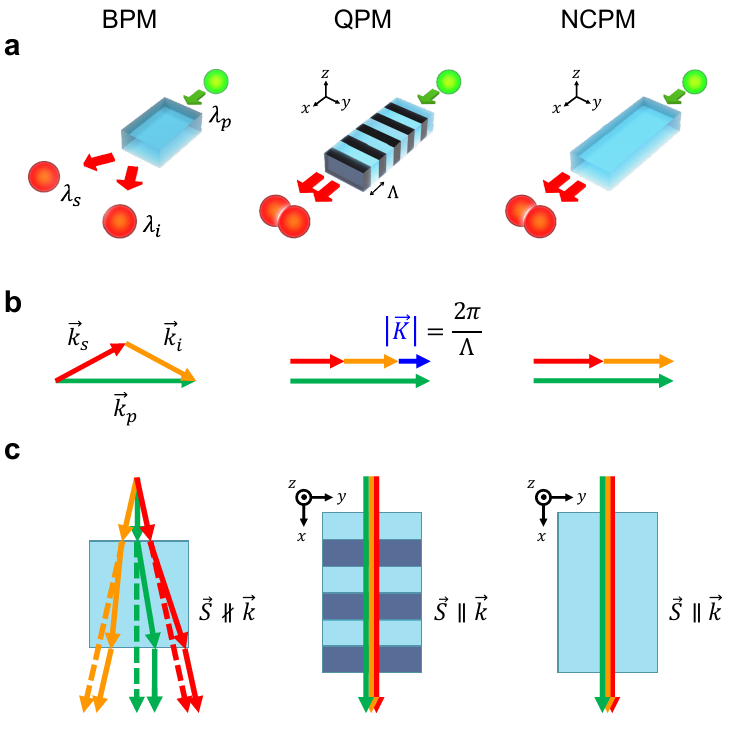}
\caption{\textbf{Various types of the SPDC.} \textbf{a,} Schematic illustration of the SPDC with birefringent phase matching (BPM), quasi phase matching (QPM), and non-critical phase matching (NCPM). In each case, signal and idler photons are generated at wavelengths of $\lambda_{s}$ and $\lambda_{i}$, respectively, by pumping the crystal with a pump photon at a wavelength of $\lambda_{p}$. The black arrows indicate one of the optical axes of the crystal. In QPM, the direction of ferroelectric polarization of the crystal is periodically inverted with the poling period $\Lambda$. \textbf{b,} Graphical description of PM conditions for SPDC with different PM techniques. $\Vec{k}_{p}$, $\Vec{k}_{s}$, and $\Vec{k}_{i}$ represent the wave vector of interacting pump, signal, and idler photons, respectively. $\Vec{K}$ in QPM is the grating vector introduced by the periodic structure in the crystal. \textbf{c,} Spatial walk-offs among interacting photons in SPDC process with different PM techniques. The solid and dashed arrows refer to the directions of the Poynting vectors and wave vectors of the interacting photons, respectively. Spatial walk-off between the Poynting vector $\Vec{S}$ and the wave vector $\Vec{k}$ of a photon occurs when the photon does not propagate along the optical axis of the crystal.}
\label{fig1}
\end{figure}

Let us begin by comparing three different types of PMs: BPM, QPM, and NCPM. Figure~\ref{fig1}\textbf{a} shows schematics of the SPDC processes using different PM techniques in a nonlinear crystal and the direction of the optical axes ($x,y,z$) of the crystal for each case. BPM, also called critical phase matching, is a technique that aligns interacting photons at specific angles to each other and the optical axes. The directions of their wave vectors $\Vec{k}$ are shown in Fig.~\ref{fig1}\textbf{b}. The BPM condition can be expressed by

\begin{equation}
\Delta \Vec{k}=\Vec{k}_{p}-\Vec{k}_{s}-\Vec{k}_{i}=0,
\label{eq1}
\end{equation}

\noindent where $p$, $s$, and $i$ correspond to pump, signal, and idler photons, respectively. The direction of BPM also affects the photon generation rate in SPDC ($\eta_{\text{SPDC}}$), which is given by

\begin{equation}
\eta_{\text{SPDC}}\propto d_{\text{eff}}^{2} L^{2} \text{sinc}^{2}(\frac{\Delta k L}{2}).
\label{photon_generation_rate}
\end{equation}

\noindent Note that $\eta_{\text{SPDC}}$ is proportional to the square of the effective nonlinear optical coefficient ($d_{\text{eff}}$) and the interaction length $L$. For BPM, $d_{\text{eff}}$ is expressed as a function of the polar and azimuthal angles ($\theta$ and $\phi$) defining the direction of the pump. Typically, the BPM direction is not parallel to the optical axes, which prevents the full exploitation of the nonlinearity of the crystal. For example, in a BBO crystal, $d_{\text{eff}}$ is expressed as $d_{\text{22}}$cos$^2$($\theta+\rho$)$\cos{3\phi}$, where $\rho$ denotes the walk-off angle.

Additionally, for BPM, spatial walk-off reduces $\eta_{\text{SPDC}}$. When the photon propagates in a direction that is not parallel to one of the optical axes of the nonlinear crystal, the Poynting vectors $\Vec{S}$ of interacting photons become separated, disrupting their spatial overlap. This limits the interaction length $L$ within the crystal and thereby decreases the efficiency of SPDC.

QPM is a technique that satisfies the PM condition in the direction parallel to one of the optical axes of a crystal, which corresponds to the propagating direction of all interacting photons. As shown in Fig.~\ref{fig1}\textbf{a}, QPM uses a PP crystal that is obtained by inverting the ferroelectric domains of the crystal periodically. The PP structure introduces a grating vector of $|\Vec{K}| = 2\pi$/$\Lambda$ to compensate for phase mismatch, allowing us to realize a collinear SPDC process in the direction parallel to one of the optical axes. The QPM condition can be expressed by

\begin{equation}
\Delta k=|\Vec{k}_{p}-\Vec{k}_{s}-\Vec{k}_{i}-\Vec{K}|=0,
\label{eq2}
\end{equation}

\noindent as shown in Fig.~\ref{fig1}\textbf{b}. In QPM, $d_{\text{eff}}$ is linearly proportional to a single nonlinear optical coefficient associated with the optical axis along which all the interacting photons propagate, and it is not affected by the angular dependence. In addition, there is no spatial walk-off in QPM, which allows a high $\eta_{\text{SPDC}}$ to be achieved by increasing the crystal length $L$.

However, a theoretical limit still exists in QPM for the value of $d_{\text{eff}}$. Due to the PP structure in the crystal, the nonlinear optical coefficient $d(l)$ exhibits spatial dependence as interacting photons propagate within the PP crystal. $d(l)$ in a PP crystal for QPM is given by


\begin{eqnarray}
d(l)&=&d_{\text{p}}\text{sign}[\text{cos}(2\pi l/\Lambda)] \nonumber \\ 
&=&d_{\text{p}}\sum_{m=-\infty}^{\infty}~G_{m}\text{exp}(ik_{m}l)
\label{d_dependence}
\end{eqnarray}
where $d_{\text{p}}$ is the nonlinear optical coefficient associated with the QPM direction and the specific type of interaction. $G_{m}=\frac{2}{m\pi} \text{sin}(\frac{m\pi}{2})$, the coefficient of the $m$-th grating order in the PP structure, is given by Fourier series expansion of $d(l)$~\cite{boyd}. Consequently, $d_{\text{eff}}$ in $m$-th order QPM is given by

\begin{equation}
d_{\text{eff}}=d_{\text{p}}G_{m}. 
\label{d_qpm}
\end{equation}
For first-order QPM, $d_{\text{eff}}$ decreases by a factor of $2/\pi$ and the corresponding $\eta_{\text{SPDC}}$ is reduced to $(2/\pi )^2 \approx40\%$ compared to the case using $d_{\text{p}}$.

NCPM is a PM technique that enables a collinear SPDC process with no spatial walk-off, without requiring a PP structure in the nonlinear crystal. It is illustrated in Fig.~\ref{fig1}\textbf{a}. In NCPM, similar to QPM, all interacting photons propagate along the same direction, which corresponds to one of the optical axes of the bulk nonlinear crystal. This leads to the absence of spatial walk-off in NCPM, allowing for a high $\eta_{\text{SPDC}}$, as shown in Fig.~\ref{fig1}\textbf{c}. Note that NCPM does not use a grating vector for phase matching from its utilization of a bulk nonlinear crystal. The Type-\rom{2} NCPM condition can be expressed by

\begin{equation}
\Delta k=\left| \Vec{k}_{p}^{(\alpha)}-\Vec{k}_{s}^{(\beta)}-\Vec{k}_{i}^{(\gamma)} \right|=0,
\label{eq7}
\end{equation}
where the superscripts $\alpha$, $\beta$ and $\gamma$ refer to the polarization directions of the interacting photons corresponding to one of the optical axes ($x,y,z$) of a nonlinear crystal, respectively. These directions are determined by the specific interaction type and the NCPM direction. As shown in Fig.~\ref{fig1}\textbf{b}, without the grating vector $\Vec{K}$, NCPM satisfies phase matching and there is no reduction in $d_{\text{p}}$. This implies that $d_{\text{eff}}$ in NCPM is $\pi$/2 times larger than in the QPM case, and the corresponding $\eta_{\text{SPDC}}$ can be enhanced by approximately 2.5 times. We also provided detailed calculations on angular tolerance and spatial walk-off, as well as the joint spectral analysis of down-converted photon pairs under NCPM, in Supplementary Note \rom{1}-C and D, respectively.

\begin{figure}[t]
\centering\includegraphics[scale=0.7]{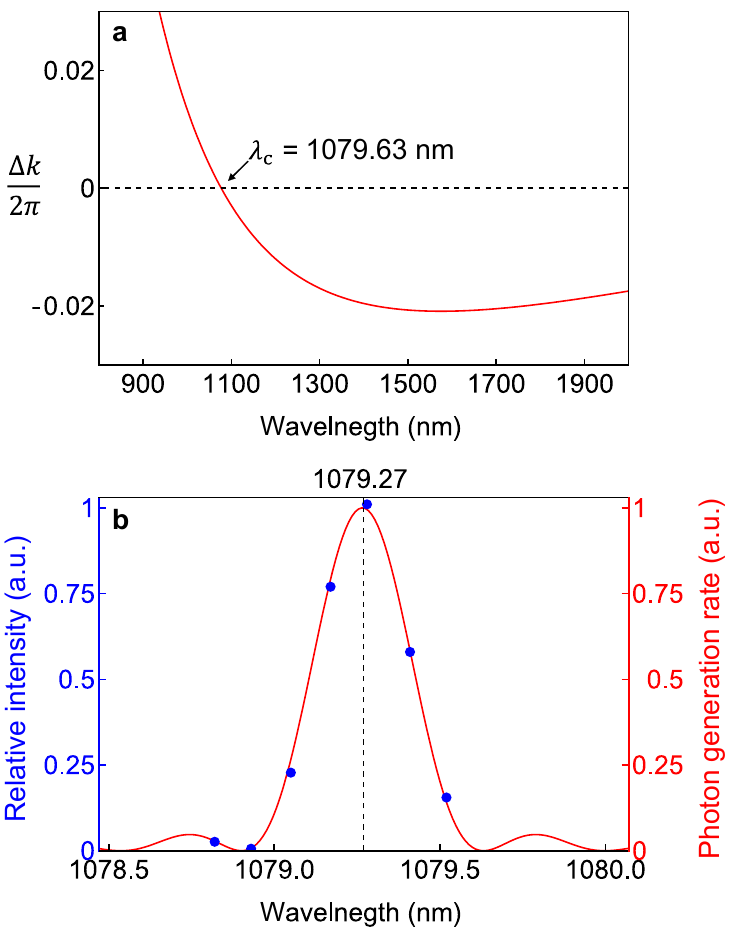}
\caption{\textbf{Phase mismatch and spectra of the generated photons in a KTP crystal with Type-\rom{2} NCPM.} \textbf{a}, Phase mismatch plotted as a function of the wavelength of the down-converted photon in frequency-degenerate Type-\rom{2} SPDC with NCPM. The NCPM condition is satisfied at a wavelength of $\lambda_{c}=1079.63$ nm. \textbf{b}, Measured normalized intensity of the SH beam (blue dot) and the calculated spectra of the normalized photon generation rate for Type-\rom{2} NCPM in a 20-mm-long bulk KTP crystal (red solid line). With a slight shift of 0.36 nm from the expected value, the expected value and the experimental result overlap at 1079.27 nm. The measured full-width-half-maximum (FWHM) value of 0.318 nm corresponds to the spectral bandwidth of the generated photons.}
	\label{fig2}
\end{figure}

\begin{figure*}[t!]
	\centering\includegraphics[scale=0.7]{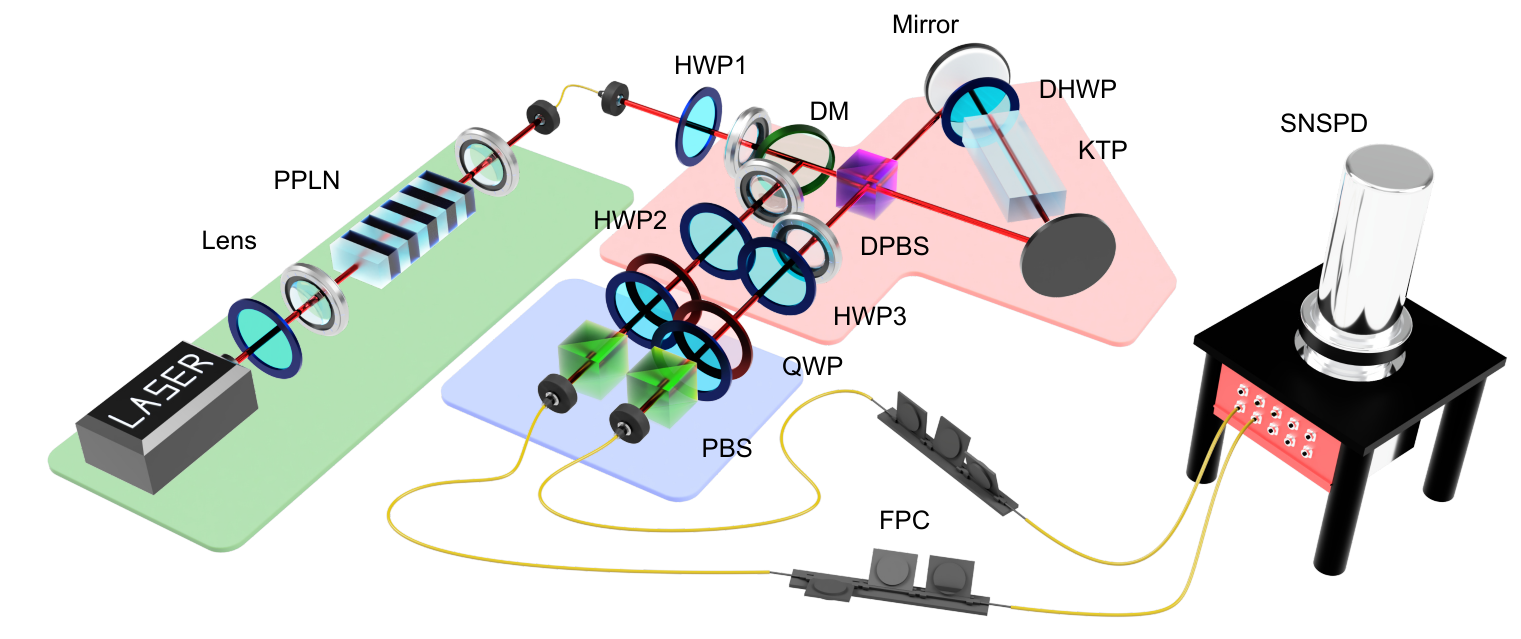}
	\caption{\textbf{Experimental setup for generating polarization-entangled photons via the frequency-degenerate collinear SPDC process with Type-\rom{2} NCPM.} Our setup consists of three parts: preparation of the pump beam (green), generation of the polarization-entangled photons (red), and two-photon polarization measurement (blue). HWP, half-wave plate; QWP, quater-wave plate; DM, dichroic mirror; DPBS, dual wavelength polarization beam splitter; DHWP, dual wavelength half-wave plate; FPC, fiber polarization controller; SNSPD, superconducting nanowire single photon detector.}
	\label{fig3}
\end{figure*}

We investigated the characteristics of the Type-\rom{2} NCPM in a bulk KTP crystal. We set the temperature of the KTP to 30$\degreecelsius$ and considered a frequency-degenerate $yyz$ NCPM interaction. See Supplementary Note~\rom{1}-A for more information on the selection of the NCPM interaction type in a KTP crystal. Figure~\ref{fig2}\textbf{a} shows the phase mismatch $\Delta k/2\pi$ in the NCPM condition as a function of the wavelength of the generated photon pairs. We used Eq.~(\ref{eq7}) to calculate $\Delta k/2\pi$, whose subscripts $\alpha, \beta,$ and $\gamma$ are set to $y, y,$ and $z$, respectively. For each wave vector $k$, the refractive indices of the KTP crystal were obtained using the Sellmeier equations from Ref.~\cite{Emanueli2003}. The NCPM condition is satisfied at a wavelength of $\lambda_{c} =$ 1079.63 nm.

In order to confirm that the SPDC process with Type-\rom{2} NCPM in a bulk KTP crystal can produce frequency-degenerate photon pairs, we used second harmonic generation (SHG) process, which is the reverse process of SPDC~\cite{Couteau2018}. Figure~\ref{fig2}\textbf{b} shows the normalized intensity of the measured SH beam while we varied the wavelength of the laser pumping a 20-mm-long bulk KTP crystal (blue dots). We also plotted the spectra of the expected $\eta_{\text{SPDC}}$ as a function of the wavelength of the generated photon pairs corresponding to the operating wavelength of the laser in Fig.~\ref{fig2}\textbf{b} (red solid line). Note that the experimental results are very well matched to the theoretical calculations, and they correspond to  the normalized $\eta_{\text{SPDC}}$ in the Type-\rom{2} NCPM. This result confirms that frequency-degenerate photon pairs can be generated at 1079.27 nm using the SPDC process with the Type-\rom{2} NCPM technique in a bulk KTP crystal. See Supplementary Note \rom{1}-B for detailed information on the calculation of the spectral bandwidth of the generated photon pairs and Supplementary Note \rom{2}-A for detailed information on our SHG measurement, respectively.

Using the PSI configuration with a 20-mm-long bulk KTP crystal, we generated frequency-degenerate polarization-entangled photon pairs at 1079.27 nm. Figure~\ref{fig3} shows the schematic diagram of our experimental setup. To prepare the pump beam for SPDC, we used a Type-0 SHG process in periodically poled lithium niobate (PPLN) and a laser operating at 1079.27 nm.  After filtering the spatial mode of the pump beam using a single-mode-fiber (SMF) and orienting its polarization state as $|D\rangle$ by half-wave plate (HWP1), the pump beam propagates in the PSI with a bulk KTP crystal. The PSI consists of a dual wavelength polarization beam splitter (DPBS), a dual wavelength half-wave plate (DHWP) oriented at 45\degree, and two mirrors~\cite{Kim2006, Fedrizzi2007}. The bulk KTP crystal is positioned at the center of the PSI. The pump beam is divided at the DPBS into horizontal ($|H\rangle$) and vertical ($|V\rangle$) components. Then, the Type-\rom{2} SPDC process with NCPM occurs bidirectionally in the bulk KTP crystal. After the generated photon pairs are combined at the DPBS, the polarization-entangled two-photon state is produced as follows:

\begin{equation}
|\Psi\rangle = \frac{1}{\sqrt{2}}(|HV\rangle+e^{i\varphi}|VH\rangle),
\label{eq8}
\end{equation}

\noindent where $\varphi$ denotes the relative phase between two states. We can prepare four maximally-entangled Bell states: $|\phi^{\pm}\rangle=(|HH\rangle\pm|VV\rangle)/\sqrt{2}$ and $|\psi^{\pm}\rangle=(|HV\rangle\pm|VH\rangle)/\sqrt{2}$ using a set of HWPs. Finally, the generated polarization-entangled states are analyzed by quantum state tomography (QST) using a set of QWP, HWP, and PBS. We used superconducting nanowire single photon detectors (SNSPDs) for detecting $\sim 1080$ nm photons. Note that the estimated detection efficiency of our SNSPDs at $\sim 1080$ nm is only $\sim 40\%$, since they  are originally designed for 1550 nm photons. See Supplementary Note \rom{2}-B for detailed information on our experimental setup.

\begin{figure}[t]
	\centering
        \includegraphics[scale=0.7]{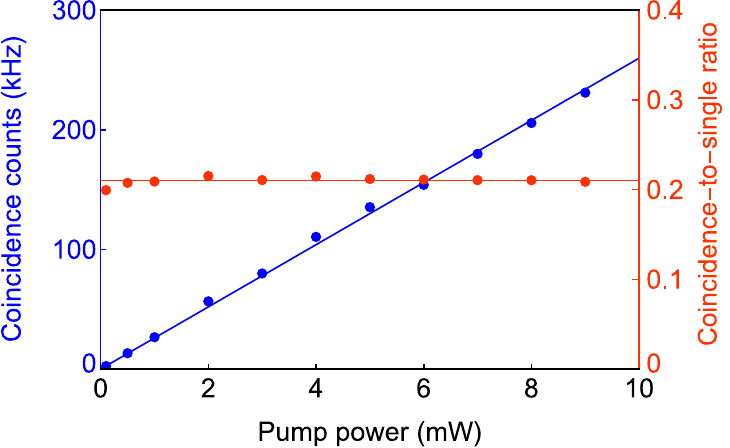}
	\caption{\textbf{Photon pair generation rate (blue dots) and the corresponding coincidence-to-single-photon ratio (red dots).} The experimental data points were obtained as the power of the SH beam varied. The photon pair generation rate was obtained by pumping a bulk KTP crystal with the horizontally polarized SH beam. The fitted curves show that the brightness of the generated photon pairs and the corresponding coincidence-to-single-photon ratio remain at  25.4 kHz/mW and 21\%, respectively, regardless of the pump power.  The error bars are too small to be visible.}
	\label{fig4}
\end{figure}

We observed a brightness of 25.1 $\pm$ 0.158 kHz/mW for the generated polarization-entangled photon pairs, with a coincidence-to-single-photon ratio $\eta_{c}=\frac{S_{c}}{\sqrt{S_{s}S_{i}}} = 21\%$, where $S_c, S_s,$ and $S_i$ refer to the count rates of the measured coincidences, signal photons, and idler photons, respectively. The errors corresponding to one standard deviation were estimated assuming Poissonian statistics. Figure~\ref{fig4} shows the photon pair generation rates and the corresponding coincidence-to-single-photon ratio. Note that the coincidence-to-single-photon ratio remains consistent as we vary the power of the pump beam. These results demonstrate that we have realized an ultra-bright polarization-entangled photon source. Table~\ref{tabcomp} compares of our work with previously reported C-band PPKTP sources based on Type-\rom{2} SPDC~\cite{Evans2010,Bruno2014,Weston2016,Chen2017_2}. We evaluated the performance of these sources using a parameter $\mathcal{U}$, which represents the brightness per unit crystal length and pump power, assuming no detection loss~\cite{boyd,Fiorentino2007}. Without losing spectral purity of photons, we achieved ultra-bright polarization-entangled photon pairs using the PSI configuration with a bulk KTP crystal.


We then measured the Hong-Ou-Mandel (HOM) interference to verify the indistinguishability of the generated photon pairs, confirming that the generated photons are frequency-degenerate. For  photon pairs generated from a horizontally (vertically)-polarized pump photon, we obtained HOM visibility of 0.999$\pm$0.002 (0.990$\pm$0.004) without using spectral filters on the down-converted photons. The measured HOM interference fringes are provided in Supplementary Note \rom{2}-C.

\begin{figure}[t]
\centering
\includegraphics[scale=0.7]{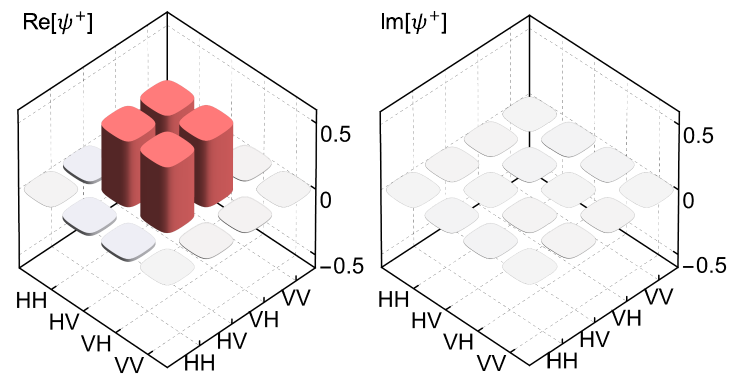}
\caption{\textbf{Reconstructed density matrix for the Bell state $|{\psi}^{+}\rangle$.} The real part (left) and the imaginary part (right) of the density matrix are shown.}
\label{fig5}
\end{figure}

We prepared four maximally-entangled Bell states $|\psi^{\pm}\rangle$ and $|\phi^{\pm}\rangle$ and performed QST to estimate the qualities of the prepared polarization-entangled Bell states. Figure~\ref{fig5} shows the real and imaginary parts of the reconstructed density matrix for one of the prepared Bell states, $|\psi^+\rangle$. The density matrices for the other Bell states are provided in Supplementary Note \rom{2}-D. We estimated the reconstructed matrices using three parameters: purity ($\mathcal{P}$), concurrence ($\mathcal{C}$), and fidelity ($\mathcal{F}$)~\cite{James2001}. Table~\ref{tab1} summarizes our experimental results for the prepared four Bell states, confirming that we successfully generated maximally-entangled states using the Type-\rom{2} SPDC process with NCPM in a bulk KTP crystal.

\begin{table*}[t]
\begin{tabular}{c|c|c|c|c|c|c|c}

                          & \textbf{\begin{tabular}[c]{@{}c@{}}Wavelength\\ (nm)\end{tabular}} & \textbf{\begin{tabular}[c]{@{}c@{}}Crystal\\ length\\ (mm)\end{tabular}} & \textbf{\begin{tabular}[c]{@{}c@{}}Detection\\ efficiency\\ (\%)\end{tabular}} & \textbf{\begin{tabular}[c]{@{}c@{}}Heralding\\ efficiency\\ (\%)\end{tabular}} & \textbf{\begin{tabular}[c]{@{}c@{}}Brightness\\ (kHz/mW)\end{tabular}} & \textbf{\begin{tabular}[c]{@{}c@{}}HOM\\ visibility\\ (\%)\end{tabular}} & \textbf{\begin{tabular}[c]{@{}c@{}}$\mathcal{U}$\\ (kHz/mW/mm)\end{tabular}} \\ \hline\hline

\textbf{Evans2010}\cite{Evans2010} & 1552 & 20 & 19 & 1.9 & 0.045 & 94.7 &0.0399 \\
\textbf{Bruno2014}\cite{Bruno2014} & 1544 & 30 & 80 & - & 4 & 91 &0.208  \\
\textbf{Weston2016}\cite{Weston2016}  & 1570 & 2 & 80 & 64 & 0.57 & $\sim$100 &0.445    \\
\textbf{Chen2017}\cite{Chen2017_2} & 1582 & 18 & 36 & 29 & 1 & 96 &0.429  \\
\textbf{our work} & 1079 & 20 & 40 & 21 & 25 & 99 & 7.81    \\
\end{tabular}
\caption{{\bf Comparison of various schemes for generating polarization-entangled photons using a Type-\rom{2} PPKTP crystal and our scheme.}}
\label{tabcomp}
\end{table*}

\begin{table}[t!]
	\centering
	\begin{tabular}{P{3em}P{7em}P{7em}P{7em}}
		 & {$\mathcal{P}$} & {$\mathcal{C}$}  & {$\mathcal{F}$} \\ \hline\hline
		$|\phi^{+}\rangle$ & 0.9887$\pm$0.0006      & 0.9869$\pm$0.0006           & 0.9888$\pm$0.0003        \\ \hline
		$|\phi^{-}\rangle$ & 0.9922$\pm$0.0007      & 0.9906$\pm$0.0006           & 0.9863$\pm$0.0003        \\ \hline
		$|\psi^{+}\rangle$ & 0.9932$\pm$0.0006      & 0.9930$\pm$0.0006           & 0.9956$\pm$0.0003        \\ \hline
		$|\psi^{-}\rangle$ & 0.9930$\pm$0.0005      & 0.9915$\pm$0.0005           & 0.9925$\pm$0.0003        \\
	\end{tabular}
        \caption{\textbf{Properties of the prepared four Bell states.} The errors correspond to one standard deviation. {$\mathcal{P}$}, purity; {$\mathcal{C}$}, concurrence; {$\mathcal{F}$}, fidelity.}
        \label{tab1}
\end{table}





In summary, we investigated the conditions for generating frequency-degenerate photon pairs via the SPDC process with Type-\rom{2} NCPM in a bulk KTP crystal and experimentally demonstrated their polarization-entanglement using the PSI configuration. Compared to BPM and QPM, our polarization-entangled photon sources using Type-\rom{2} NCPM have several advantages. First, since NCPM does not use an additional grating vector for phase matching, there is no reduction in $d_{{\rm eff}}$, allowing for a 2.5-fold enhancement in brightness compared to QPM. Moreover, NCPM does not require a PP structure, thus avoiding the need for intricate post-processing steps such as PP, which incurs high production costs and limits the crystal thickness. Additionally, all remaining advantages of QPM over BPM, such as higher brightness and the absence of spatial walk-off, still apply to NCPM. The generated photon source exhibits an ultra-brightness of 25.1 $\pm$ 0.158 kHz/mW and high spectral purity, corresponding to an HOM visibility of 0.99. The quality of the prepared entangled state is confirmed by a concurrence $\mathcal{C}\geq 0.9869$ and the state fidelity $\mathcal{F}\geq 0.9863$.

The generated polarization-entangled photon pairs are at 1079.27 nm, which falls within the optical communication T-band (1000--1260 nm). Due to its high transparency in the atmospheric environment~\cite{plank2012,Guan2022}, this region has found wide applications, such as satellite-to-Earth communication~\cite{xue2016,Courde2017,Tang2018} and free-space light detection and ranging (LIDAR) sensing systems~\cite{zhu2017,Vaughan2019,Li2023}. Additionally, the T-band offers a high channel density of 62 THz~\cite{Tsuda2016}, making it a potential solution to the capacity crunch in fiber networks~\cite{richardson2010,Idris2016}.  

The NCPM technique can be applied not only to  KTP crystals but also to other nonlinear crystals in the SPDC process. The NCPM technique does not restrict the crystal design for SPDC, allowing us to prepare a bright photon source using a longer crystal and enabling high-dimensional entanglement with a larger crystal surface~\cite{Hu2021, Slussarenko2022, Hu2020, Li2020}. Furthermore, it can be applied to the fabrication of waveguide SPDC photon sources~\cite{Wang2021}. We believe that our results provide a new methodology for generating efficient quantum resources, contributing to the development of photonic quantum information technology.

\end{document}